# Mathematical model of inverted analemmatic sundial on the example of Belogorsk sundial of the Bronze Age

**Larisa N. Vodolazhskaya**

Institute of Physics and Technology, V.I. Vernadsky Crimean Federal University, Simferopol, Republic of Crimea, Russian Federation
Southern Federal University, Rostov-on-Don, Russian Federation
E-mails: larisavodol@aaatec.org, larisavodol@gmail.com

**Abstract**

This article proposes a mathematical model of an inverted analemmatic sundial, provides formulas for calculating the coordinates of their hour markers. On the example of the Belogorsk sundial, a method is described for determining the accuracy of time measurement using an inverted analemmatic sundial, in which cup marks are used as hour markers. The obtained value of the accuracy of time measurement with the help of the Belogorsk sundial is 5÷6 minutes and is quite good for a measuring instrument of the Bronze Age, and is also higher than the accuracy of time measurement with the help of a water clock. As a result, it is concluded that the hypothesis is confirmed that the invention of the sundial was associated with the need to improve the accuracy of time measurement using a water clock.

In 2017 the expedition of the Museum-Reserve "Scythian Neapolis" in the process of archaeological excavations of kurgan 1 of the kurgan grave field Prolom II (1,7 km east of the Prolom village), located in the Belogorsk district of Crimea. A break, among the blockage of the stone shell of the kurgan, a stone with hollowed cup marks was found (Fig. 1)[1].

The main feature of the slab is small cup marks densely located along the edge of the slab and partially filling its inner area. On one side of the slab along its edge, cup marks are applied, forming a curved line resembling the arc of an ellipse.

In the process of research, it was found that the Belogorsk slab is a sundial of about the 14th-13th centuries BC and belongs to the Srubnaya culture. By type of sundial, it is closest to the analemmatic sundial. However, the principle of the hourly markings of the Belogorsk slab is so unique that it made it possible to distinguish a new type of sundial - an inverted analemmatic sundial. Unlike typical analemmatic sundials, the gnomon remains motionless throughout the year, and in accordance with the analemma, the "dial" moves - an ellipse of hour markers consisting of cup marks, i.e. gnomon and hour markers (cup marks) change places in terms of mobility. The movement of hour marks is not literal, but is replaced by several rows of cup marks, which are fragments of ellipses of hour marks for different months of the year (Vodolazhskaya, 2022, 1-29).

---

[1] Museum inventory number KM 7265/5 (Historical and Archaeological Museum-Reserve "Scythian Neapolis").

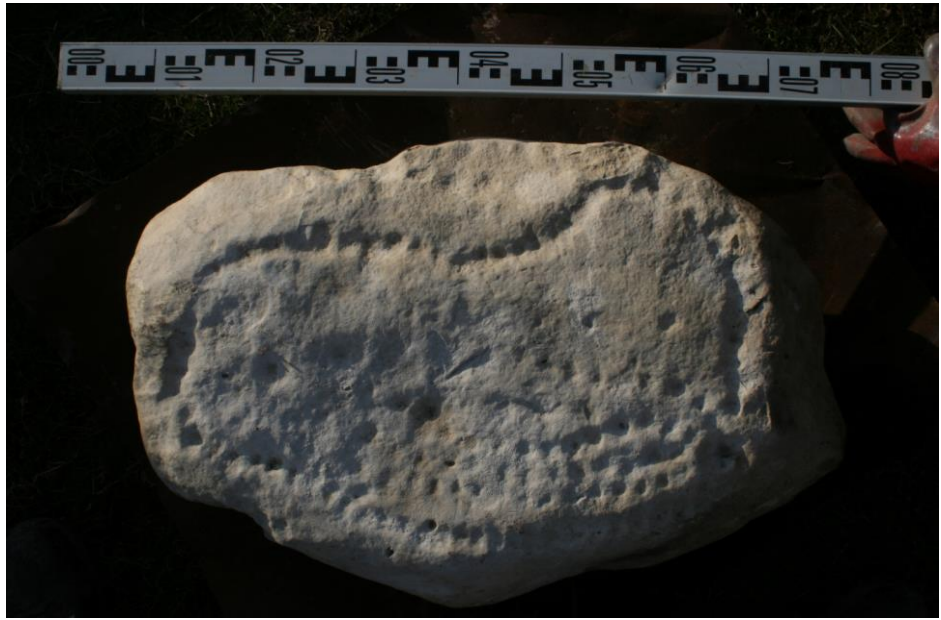

**Figure 1.** Kurgan grave field Prolom II, kurgan 1, slab with cup marks (photo by V. Nuzhdenko, 2017).

The construction of the hour markers of the inverted sundial in the previous study was done using geometric transformations. The points of the gnomon analemma were successively reflected about the X-axis and the Y-axis. Then, each hour mark of the analemmatic sundial was replaced by a set of hour markers - points of the mirror-reflected analemma. In this case, the center of the analemma was located in each hour mark of the analemmatic sundial.

The mathematical model of classical analemmatic sundial is well known (Rohr, 1965, 100-106; Waugh, 1973, 108-115; Mayall, Mayall, 1994, 60-61, 186-190; Savoie, 2009, 111-124) and can be described by the following formulas:

$$M = \frac{m}{\sin \varphi} \quad (1)$$

$$x = M \cdot \sin H \quad (2)$$

$$y = M \cdot \sin \varphi \cdot \cos H \quad (3)$$

$$\text{where } H = 15° \cdot (t - 12) \ , \ t \in \ ]0; 24] \quad (4)$$

$$Z_{ws} = M \cdot tg\delta_{ws} \cdot \cos\varphi \quad (5)$$

$$Z_{ss} = M \cdot tg\delta_{ss} \cdot \cos\varphi \quad (6)$$

where $x$ – the coordinate of a point along the $X$ axis for an analemmatic sundial, $y$ – the coordinate of a point along the $Y$ axis for an analemmatic sundial, $m$ – the semi-minor axis of the ellipse, *M* – the semi-major axis of the ellipse, *φ* – the latitude of the area, *t* – the true local solar time, *H* – the hour angle of the Sun, $\delta_{ws}=-\varepsilon$ – declination of the Sun on the day of the winter solstice, $\delta_{ss}=\varepsilon$ – declination of the Sun on the day of the summer solstice, on the days of the equinox $\delta_{eq}=0$, $y=Z_{ws}$ – on the day of the winter solstice, $y=Z_{ss}$ – on the day of the summer solstice (Fig. 9).

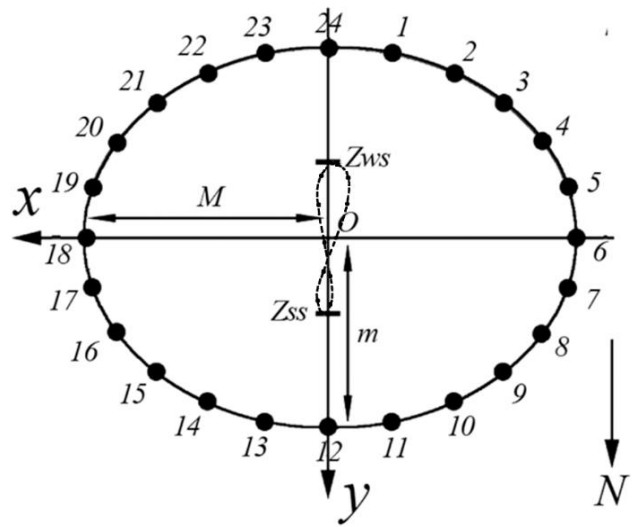

**Figure 1.** Coordinate plane with hour markers from 6 to 18 o'clock. *M* is the semi-major axis of the ellipse, *m* – the semi-minor axis of the ellipse, *O* – the center of the ellipse, $O_{ws}$ – the position of the gnomon on the winter solstice for analemmatic clock, $O_{ss}$ – the position of the gnomon on the summer solstice for the analemmatic clock. *N* – the direction to the true North.

The coordinates of the points of the analemma $Zx_n$ and $Zy_n$ of the gnomon are calculated by formulas 7 and 8:

$$Zy_n = M \cdot tg\delta_n \cdot \cos\varphi \qquad (7)$$

$$Zx_n = \frac{\eta_n}{60} \cdot M \cdot \sin(15°) \qquad (8)$$

where $Zy_n$ - the coordinates of the *n*-th point of the gnomon analemma along the *Y*-axis, $Zx_n$ - the coordinates of the *n*-th point of the gnomon analemma along the *X*-axis, $\delta_n$ - the solar declination for the date corresponding to the *n*-th point of the gnomon analemma, $\eta_n$ - the equation of time for the date corresponding to *n*-th point of the gnomon analemma; *M* - the semi-major axis of the hour markers ellipse, *φ* - the latitude of the area.

The value of the equation of time was calculated using the astronomical program HORIZONS System[2]. The calculation results for each month, starting from the day of the vernal equinox, are presented in Table 1.

As mentioned above, the location of the hour markers of the inverted sundial in the previous study was found using geometric transformations - reflection of the points of the gnomon analemma sequentially relative to the *X*-axis and the *Y*-axis, and subsequent replacement of each hour mark of the analemmatic sundial with a set of hour markers - points of reflected analemma (Fig. 2, 3).

---

[2] http://ssd.jpl.nasa.gov/?horizons

**Table 1.** The coordinates of the analemma points each month after the vernal equinox for 1500 BC, for geographic coordinates 45°04′ N, 34°36′ E: $n$ – the number of the point on the gnomon analemma, $\delta_n$ – the declination of the Sun for the date corresponding to $n$-th point of the gnomon analemma, $\eta_n$ – the time equation for the date corresponding to the $n$-th point of the gnomon analemma, $Zx_n$ – the coordinate of the $n$-th point of the gnomon analemma along the $X$ axis, $Zy_n$ – the coordinate of the $n$-th point of the gnomon analemma along the $Y$ axis (Vodolazhskaya, 2022, table 2).

| n | Date | $\delta_n$, ° | $\eta_n$, min | $Zx_n$, cm | $Zy_n$, cm |
|---|---|---|---|---|---|
| 1 | April (vernal equinox) 3 April 1500 BC | -0,05 | -7 | -0,8 | 0,0 |
| 2 | May 3 May 1500 BC | 11,13 | 5 | 0,5 | 3,5 |
| 3 | June 3 June 1500 BC | 20,07 | 10 | 1,1 | 6,6 |
| 4 | July 3 July 1500 BC | 23,83 | 6 | 0,7 | 7,9 |
| 5 | August 3 August 1500 BC | 21,20 | -2 | -0,2 | 7,0 |
| 6 | September 3 September 1500 BC | 12,67 | -2 | -0,2 | 4,0 |
| 7 | October 3 October 1500 BC | 1,12 | 4 | 0,4 | 0,4 |
| 8 | November 3 November 1500 BC | -11,22 | 9 | 1,0 | -3,6 |
| 9 | December 3 December 1500 BC | -20,37 | 6 | 0,7 | -6,7 |
| 10 | January 3 January 1500 BC | -23,82 | -8 | -0,9 | -7,9 |
| 11 | February 3 February 1500 BC | -19,98 | -19 | -2,1 | -6,5 |
| 12 | March 3 March 1500 BC | -11,72 | -18 | -2,0 | -3,7 |

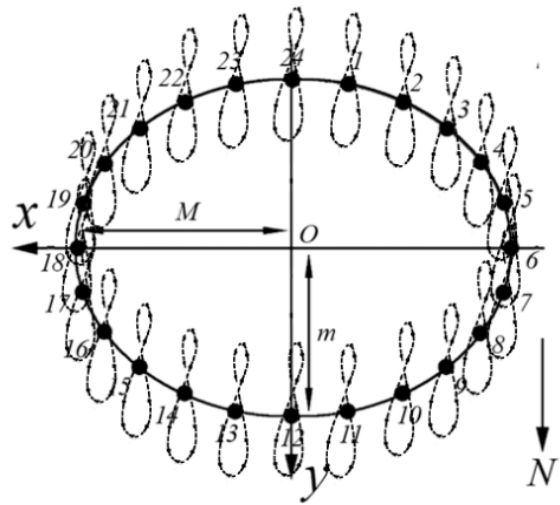

**Figure 2.** Coordinate plane with hour markers and their analemmas of an inverted analemmatic sundial. *M* – the semi-major axis of the ellipse of hour markers, *m* - the minor semi-axis of the ellipse of hour markers, *O* – the center of the ellipse. *N* – the direction to the true North.

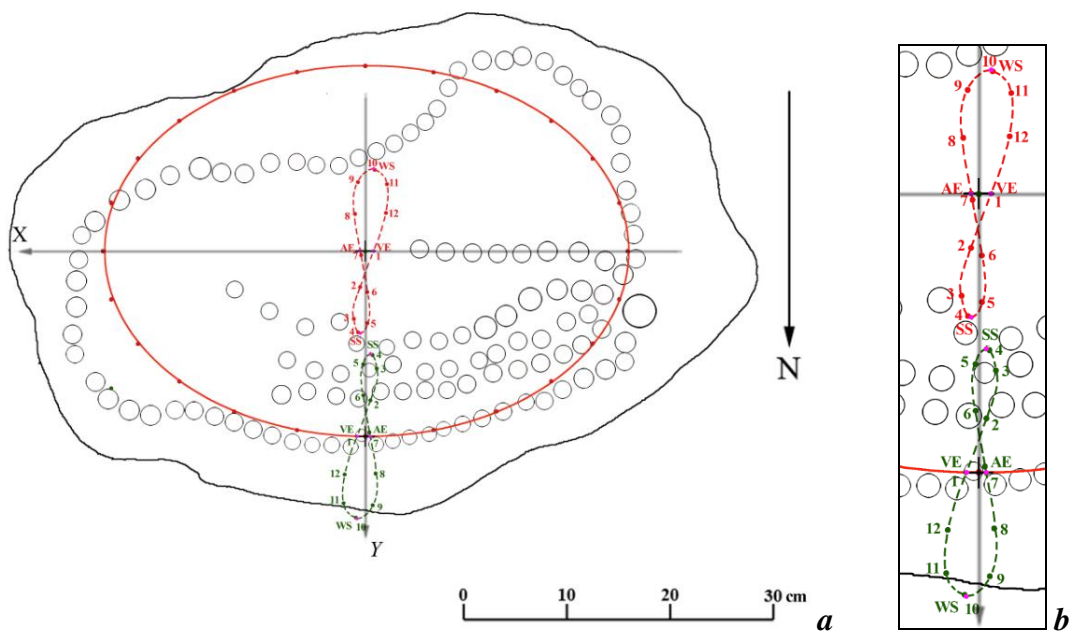

**Figure 3.** Plan-scheme of the Belogorsk slab: *a* – the surface of the slab with an analemma for a moving gnomon with the corresponding ellipse of hour marks (highlighted in red) and an analemma for cup marks corresponding to 12 hours during the year every month (highlighted in green); *b* – enlarged image of both analemmas. WS – the point at which you need to install a gnomon (red analemma) or place a cup mark (green analemma) on the day of the winter solstice, SS – on the summer solstice, AE – on the autumnal equinox, VE – on the vernal equinox. The numbers on the analemma show the points where you need to install the gnomon (red analemma) or place the cup marks 12 hours after the corresponding number of months after the vernal equinox (green analemma) (Vodolazhskaya, 2022, fig. 14)[3].

---

[3] In the figures of the previous article (Vodolazhskaya, 2022, 1-29), a typo was made - the numbering of the analemma points was shifted by one. Number one was put next to the point of the month following the spring equinox, but it was necessary to put it next to the point of spring equinox.

The mathematical model of the inverted analemmatic sundial, in this case, is described by the same formulas as the mathematical model of the analemmatic sundial, with the exception of formulas 2 and 3 for calculating the coordinates of hour marks. The formulas for calculating the coordinates of the hour markers of an inverted analemmatic sundial will be as follows:

$$x_n = M \cdot \sin H - Zx_n \qquad (9)$$

$$y_n = M \cdot \sin \varphi \cdot \cos H - Zy_n \qquad (10)$$

where $x_n$ – the coordinates of the hour markers of the inverted analemmatic sundial along the *X*-axis for the date corresponding to the *n*-th point of the gnomon analemma, $y_n$ – the coordinates of the hour markers along the *Y*-axis for the date corresponding to the *n*-th point of the gnomon analemma, $Zx_n$ – the coordinates of the *n*-th point of gnomon analemma along the *X*-axis, $Zy_n$ – the coordinates of the *n*-th point of the gnomon analemma along the *Y*-axis, $\delta_n$ – the solar declination for the date corresponding to the *n*-th point of the gnomon analemma, $\eta_n$ – the equation of time for the date corresponding to the *n*-th point of the gnomon analemma; $M$ – the semi-major axis of the hour markers ellipse, $\varphi$ – the latitude of the area.

The results of calculations using formulas 9 and 10 of the coordinates of the hour markers of the inverted analemmatic sundial, in the interval from 6 to 13 hours, corresponding to the observed rows of cup marks, for each month, starting from the day of the vernal equinox, are presented in tables 2 and 3.

**Table 2.** The coordinates of the hour markers of the inverted analemmatic sundial along the *X*-axis for 1500 BC, calculated by formula 9, for geographical coordinates 45°04′ N, 34°36′ E: *n* – the number of the point of the analemma, $x_n$ – the coordinates along the *X*-axis of the *n*-th points on the analemma of the hour marks in the interval from 6 to 13 o'clock, *t* – the time (hours).

| t \ n | 6 | 6,5 | 7 | 7,5 | 8 | 8,5 | 9 | 9,5 | 10 | 10,5 | 11 | 11,5 | 12 | 12,5 | 13 |
|---|---|---|---|---|---|---|---|---|---|---|---|---|---|---|---|
| 1 | -24,6 | -24,4 | -23,8 | -22,7 | -21,2 | -19,4 | -17,2 | -14,7 | -11,9 | -8,9 | -5,8 | -2,5 | 0,8 | 4,1 | 7,4 |
| 2 | -25,9 | -25,7 | -25,1 | -24,0 | -22,5 | -20,7 | -18,5 | -16,0 | -13,2 | -10,2 | -7,1 | -3,8 | -0,5 | 2,8 | 6,1 |
| 3 | -26,5 | -26,3 | -25,7 | -24,6 | -23,1 | -21,3 | -19,1 | -16,6 | -13,8 | -10,8 | -7,7 | -4,4 | -1,1 | 2,2 | 5,5 |
| 4 | -26,1 | -25,9 | -25,3 | -24,2 | -22,7 | -20,9 | -18,7 | -16,2 | -13,4 | -10,4 | -7,3 | -4,0 | -0,7 | 2,6 | 5,9 |
| 5 | -25,2 | -25,0 | -24,4 | -23,3 | -21,8 | -20,0 | -17,8 | -15,3 | -12,5 | -9,5 | -6,4 | -3,1 | 0,2 | 3,5 | 6,8 |
| 6 | -25,2 | -25,0 | -24,4 | -23,3 | -21,8 | -20,0 | -17,8 | -15,3 | -12,5 | -9,5 | -6,4 | -3,1 | 0,2 | 3,5 | 6,8 |
| 7 | -25,8 | -25,6 | -25,0 | -23,9 | -22,4 | -20,6 | -18,4 | -15,9 | -13,1 | -10,1 | -7,0 | -3,7 | -0,4 | 2,9 | 6,2 |
| 8 | -26,4 | -26,2 | -25,6 | -24,5 | -23,0 | -21,2 | -19,0 | -16,5 | -13,7 | -10,7 | -7,6 | -4,3 | -1,0 | 2,3 | 5,6 |
| 9 | -26,1 | -25,9 | -25,3 | -24,2 | -22,7 | -20,9 | -18,7 | -16,2 | -13,4 | -10,4 | -7,3 | -4,0 | -0,7 | 2,6 | 5,9 |
| 10 | -24,5 | -24,3 | -23,7 | -22,6 | -21,1 | -19,3 | -17,1 | -14,6 | -11,8 | -8,8 | -5,7 | -2,4 | 0,9 | 4,2 | 7,5 |
| 11 | -23,3 | -23,1 | -22,5 | -21,4 | -19,9 | -18,1 | -15,9 | -13,4 | -10,6 | -7,6 | -4,5 | -1,2 | 2,1 | 5,4 | 8,7 |
| 12 | -23,4 | -23,2 | -22,6 | -21,5 | -20,0 | -18,2 | -16,0 | -13,5 | -10,7 | -7,7 | -4,6 | -1,3 | 2,0 | 5,3 | 8,6 |

**Table 3.** The coordinates of the hour markers of the inverted analemmatic sundial along the *Y*-axis for 1500 BC, calculated by formula 9, for geographical coordinates 45°04′ N, 34°36′ E: *n* – the number of the point of the analemma, $y_n$ – the coordinates along the *Y*-axis of the *n*-th points on the analemma of the hour marks in the interval from 6 to 13 o'clock, *t* – the time (hours).

| t \ n | \ | | | | | | $y_n$ | | | | | | | |
|---|---|---|---|---|---|---|---|---|---|---|---|---|---|---|
|  | 6 | 6,5 | 7 | 7,5 | 8 | 8,5 | 9 | 9,5 | 10 | 10,5 | 11 | 11,5 | 12 | 12,5 | 13 |
| 1 | 0,0 | 2,3 | 4,7 | 6,9 | 9,0 | 11,0 | 12,7 | 14,3 | 16,6 | 15,6 | 17,8 | 17,4 | 17,8 | 18,0 | 17,4 |
| 2 | -3,5 | -1,2 | 1,2 | 3,4 | 5,5 | 7,5 | 9,2 | 10,8 | 13,1 | 12,1 | 14,3 | 13,9 | 14,3 | 14,5 | 13,9 |
| 3 | -6,6 | -4,3 | -1,9 | 0,3 | 2,4 | 4,4 | 6,1 | 7,7 | 10,0 | 9,0 | 11,2 | 10,8 | 11,2 | 11,4 | 10,8 |
| 4 | -7,9 | -5,6 | -3,2 | -1,0 | 1,1 | 3,1 | 4,8 | 6,4 | 8,7 | 7,7 | 9,9 | 9,5 | 9,9 | 10,1 | 9,5 |
| 5 | -7,0 | -4,7 | -2,3 | -0,1 | 2,0 | 4,0 | 5,7 | 7,3 | 9,6 | 8,6 | 10,8 | 10,4 | 10,8 | 11,0 | 10,4 |
| 6 | -4,0 | -1,7 | 0,7 | 2,9 | 5,0 | 7,0 | 8,7 | 10,3 | 12,6 | 11,6 | 13,8 | 13,4 | 13,8 | 14,0 | 13,4 |
| 7 | -0,4 | 1,9 | 4,3 | 6,5 | 8,6 | 10,6 | 12,3 | 13,9 | 16,2 | 15,2 | 17,4 | 17,0 | 17,4 | 17,6 | 17,0 |
| 8 | 3,6 | 5,9 | 8,3 | 10,5 | 12,6 | 14,6 | 16,3 | 17,9 | 20,2 | 19,2 | 21,4 | 21,0 | 21,4 | 21,6 | 21,0 |
| 9 | 6,7 | 9,0 | 11,4 | 13,6 | 15,7 | 17,7 | 19,4 | 21,0 | 23,3 | 22,3 | 24,5 | 24,1 | 24,5 | 24,7 | 24,1 |
| 10 | 7,9 | 10,2 | 12,6 | 14,8 | 16,9 | 18,9 | 20,6 | 22,2 | 24,5 | 23,5 | 25,7 | 25,3 | 25,7 | 25,9 | 25,3 |
| 11 | 6,5 | 8,8 | 11,2 | 13,4 | 15,5 | 17,5 | 19,2 | 20,8 | 23,1 | 22,1 | 24,3 | 23,9 | 24,3 | 24,5 | 23,9 |
| 12 | 3,7 | 6,0 | 8,4 | 10,6 | 12,7 | 14,7 | 16,4 | 18,0 | 20,3 | 19,3 | 21,5 | 21,1 | 21,5 | 21,7 | 21,1 |

Despite the fact that in recent years several slabs with the markings of analemmatic Bronze Age sundials have already been discovered (Vodolazhskaya, 2013, 68-88; Vodolazhskaya, Larenok, Nevsky, 2014, 31-53; Vodolazhskaya, Larenok, Nevsky, 2016, 96-116; Vodolazhskaya, Novichikhin, Nevsky, 2021, 73-86; Petricevic, Vodolazhskaya, 2021, 51-65), a method for assessing the accuracy of time measurement with their help has not been developed so far.

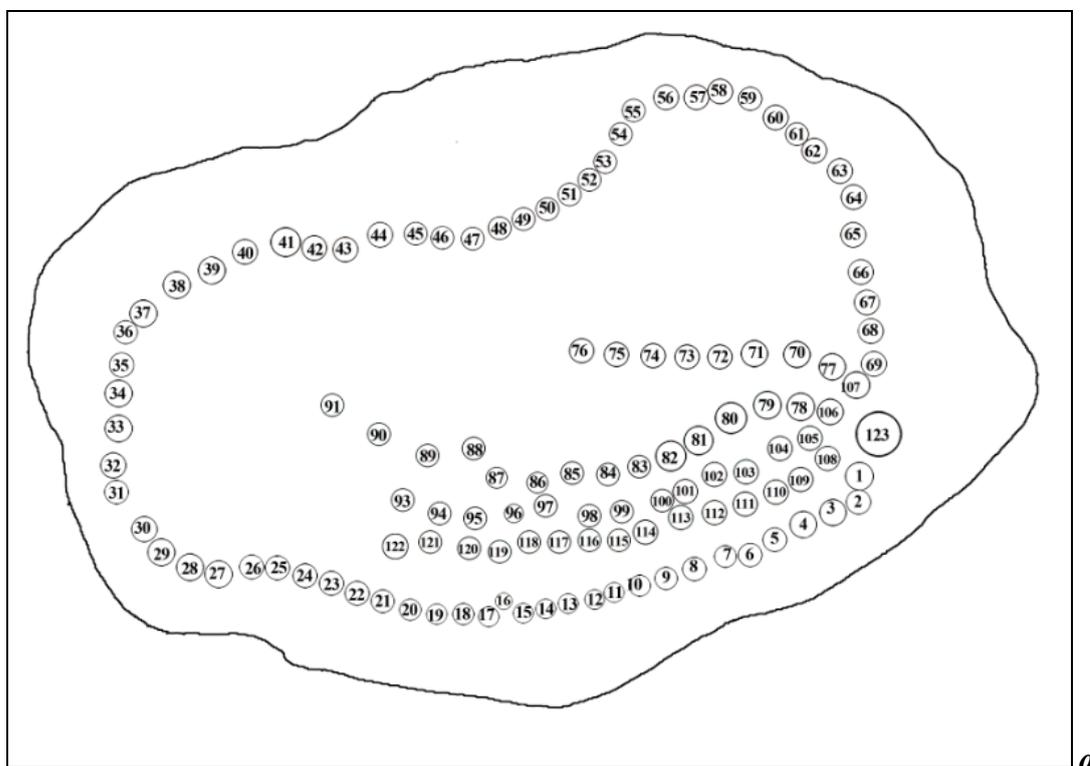

*a*

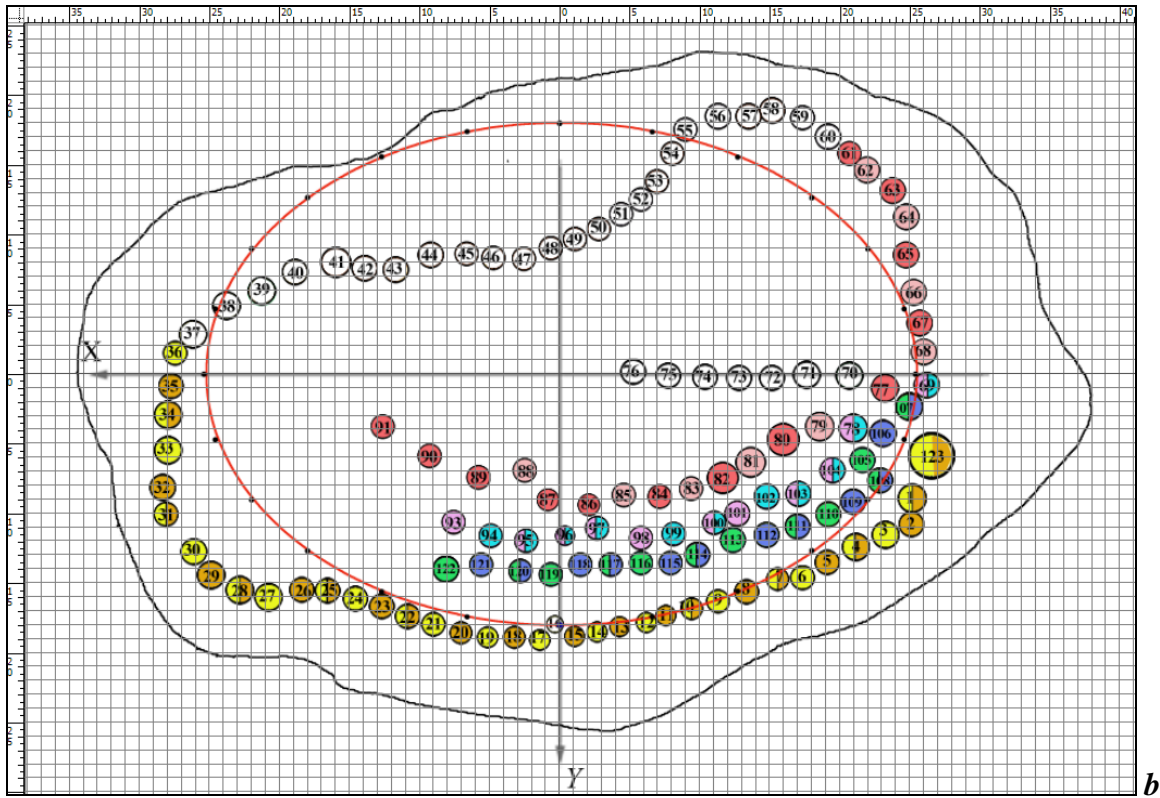

**Figure 4.** Plan-scheme of the Belogorsk slab with numbered cup marks: ***a*** – without coloring; ***b*** – with staining of rows of cup marks. The cup marks were stained depending on the correspondence to the hour marks (Fig. 5-8).

Within the framework of this study, a method is proposed for estimating the accuracy of time measurement using the Belogorsk slab with the marking of inverted analemmatic sundials, which can also be applied to slabs with the marking of analemmatic sundial in the form of cup marks. All discovered frames of analemmatic sundials of the Bronze Age are slabs with cup marks. Therefore, at the first stage, all wells on the plate were numbered (Fig. 4) (Table 4).

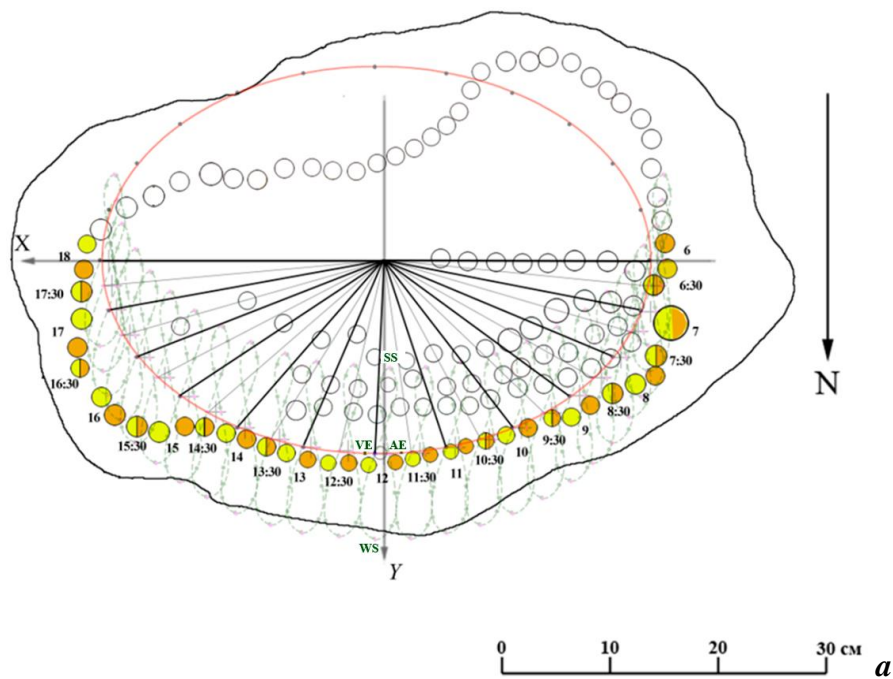

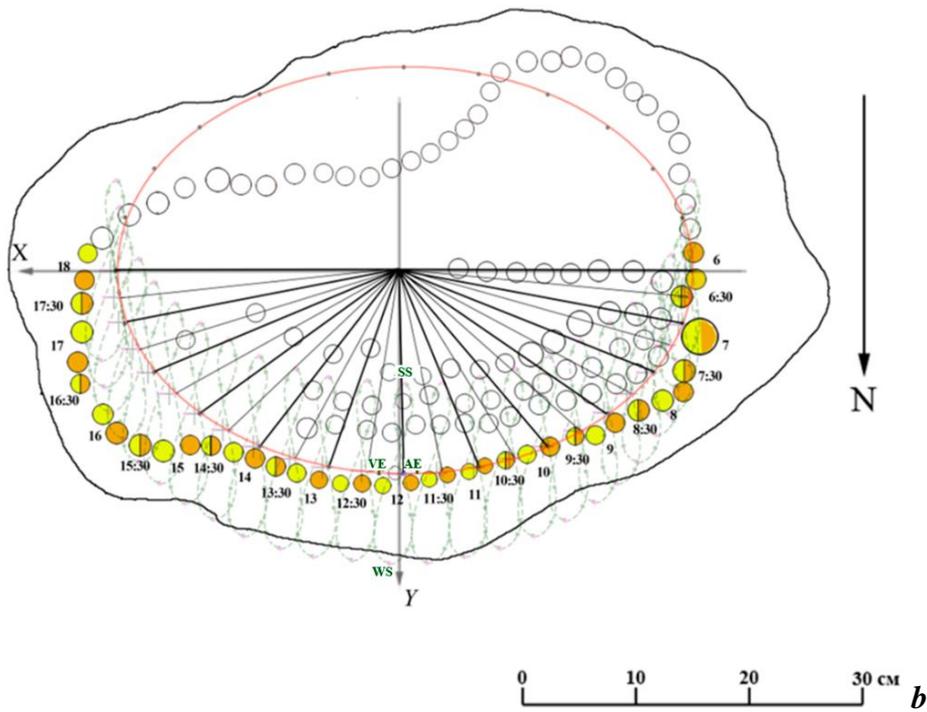

**Figure 5.** Plan-scheme of the Belogorsk slab. Hour lines and corresponding cup marks for measuring time at the equinox: *a* – vernel equinox; *b* – autumnal equinox. Cup marks for measuring time only in the vicinity of the autumnal equinox are colored orange, only in the vicinity of the vernal equinox – in yellow, for measuring time in the vicinity of both equinoxes – in both colors (Vodolazhskaya, 2022, fig. 17).

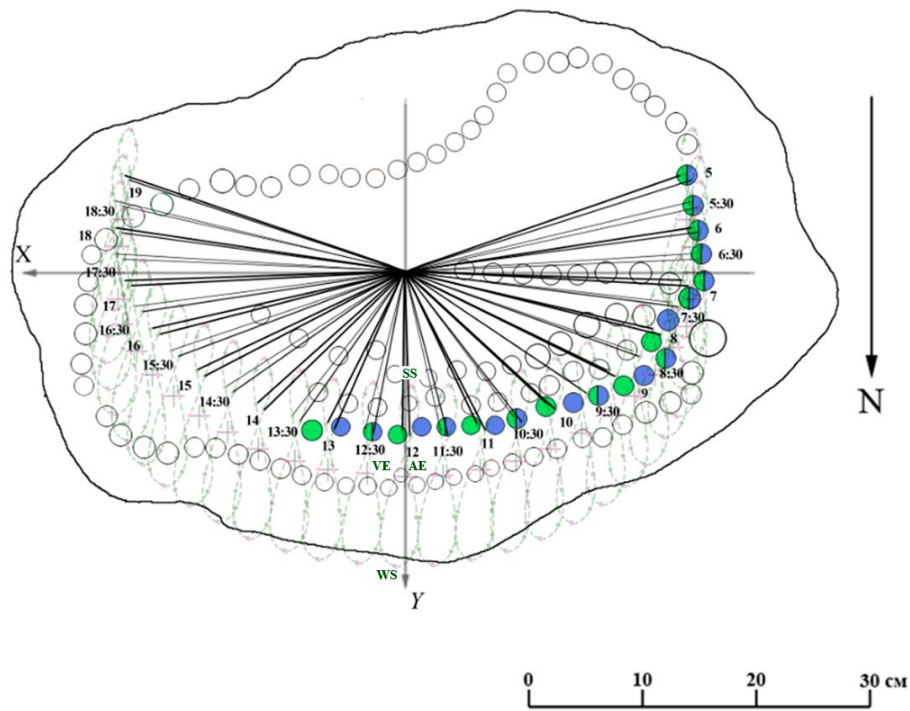

**Figure 6.** Plan-scheme of the Belogorsk slab. Hour lines and corresponding cup marks to measure the time approximately one (highlighted in blue) and five months (highlighted in green) after the vernal equinox. Cup marks for the 1st month after the VE are colored blue, for the 5th month they are green, the cup marks common for both equinoxes are in both colors (Vodolazhskaya, 2022, fig. 19).

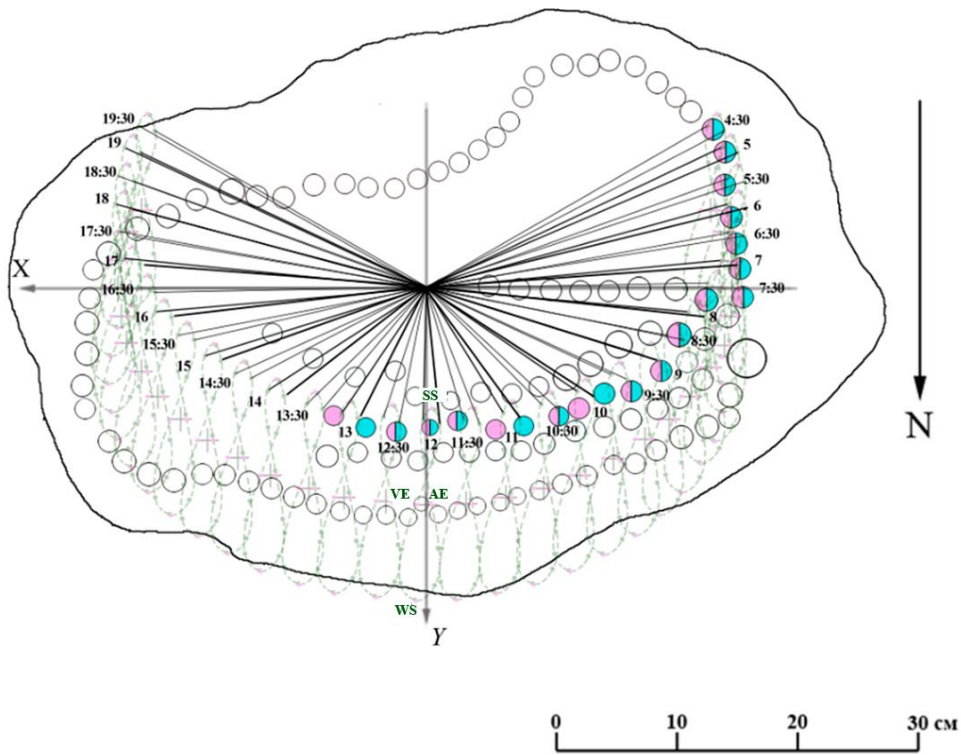

**Figure 7.** Plan-scheme of the Belogorsk slab. Hour lines and corresponding cup marks for measuring time in the period close to two months (blue) and four months (pink) after the vernal equinox. Two-color cup marks are used to measure time in both cases (Vodolazhskaya, 2022, fig. 20).

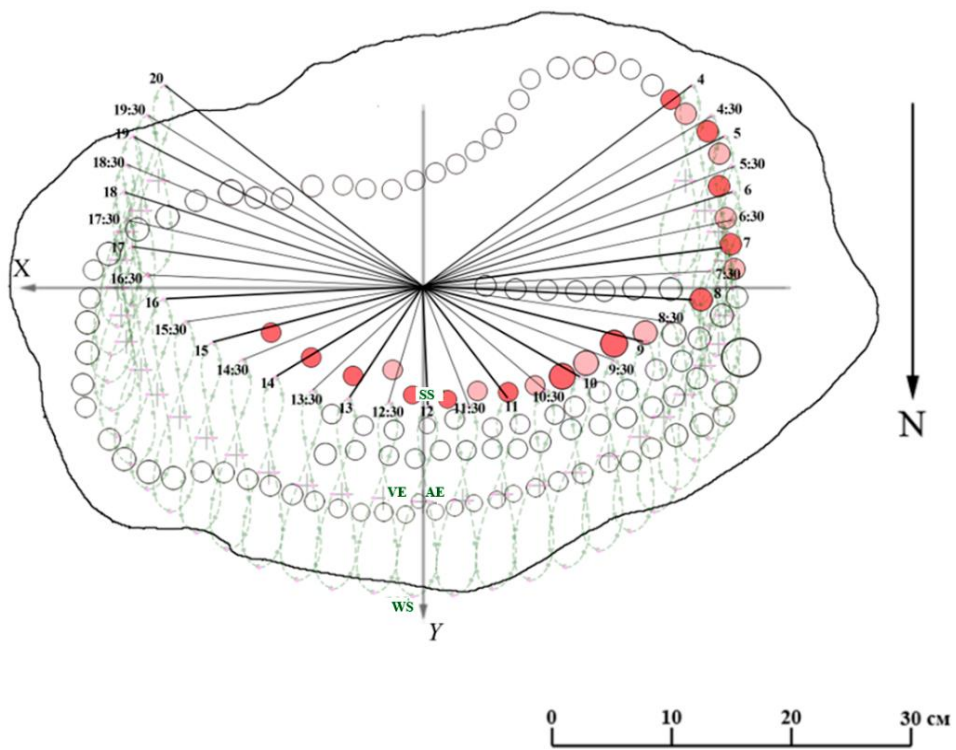

**Figure 8.** Plan-scheme of the Belogorsk slab. Hour lines and corresponding cup marks for measuring time on the days of the summer solstice. Cup marks corresponding to whole hours are highlighted in red, corresponding to halves of an hour in pale pink (Vodolazhskaya, 2022, fig. 21).

For the convenience of comparing the numbers of cup marks with the time of day in each of the considered periods (by months), we present here illustrations from the previous article (Fig. 5, 6, 7, 8) (Vodolazhskaya, 2022, 1-29).

**Table 4.** Cup marks numbers on the plan-scheme of the Belogorsk slab: $k$ – the cup mark number, $t$ – the time (hours), $n$ – the point number on the analemma.

| | \multicolumn{15}{c|}{$k$} | | | | | | | | | | | | | | |
|---|---|---|---|---|---|---|---|---|---|---|---|---|---|---|---|
| t\n | 6 | 6,5 | 7 | 7,5 | 8 | 8,5 | 9 | 9,5 | 10 | 10,5 | 11 | 11,5 | 12 | 12,5 | 13 |
| 1 | 69 | 107 | 123 | 1 | 3 | 4 | 6 | 7 | 9 | 10 | 12 | 14 | 17 | 19 | 21 |
| 7 | 68 | 107 | 123 | 1 | 2 | 4 | 5 | 7 | 8 | 10 | 11 | 13 | 15 | 18 | 20 |
| 2 | 67 | 68 | 69 | 107 | 106 | 108 | 109 | 111 | 112 | 114 | 115 | 117 | 118 | 120 | 121 |
| 6 | 67 | 68 | 69 | 107 | 105 | 108 | 110 | 111 | 113 | 114 | 116 | 117 | 119 | 120 | 122 |
| 3 | 66 | 67 | 68 | 69 | 77 | 78 | 104 | 103 | 102 | 100 | 99 | 97 | 96 | 95 | 94 |
| 5 | 66 | 67 | 68 | 69 | 77 | 78 | 104 | 103 | 101 | 100 | 98 | 97 | 96 | 95 | 93 |
| 4 | 65 | 66 | 67 | 68 | 77 | 79 | 80 | 81 | 82 | 83 | 84 | 85 | 86 | 88 | 89 |

Then the coordinates of the centers of the cup marks, corresponding to the time range from 6 to 13 hours, were measured (table 5, 6).

**Table 5.** Measured coordinates of the hour markers of the inverted analemmatic sundial along the *X*-axis: $n$ – the number of the point on the analemma, $x_{n\ meas}$ – the measured coordinates along the *X*-axis of the $n$-th points on the analemmas of the hour marks in the interval from 06:00 to 13:00, $t$ – the time (hours).

| | \multicolumn{15}{c|}{$x_{n\ meas}$} | | | | | | | | | | | | | | |
|---|---|---|---|---|---|---|---|---|---|---|---|---|---|---|---|
| t\n | 6 | 6,5 | 7 | 7,5 | 8 | 8,5 | 9 | 9,5 | 10 | 10,5 | 11 | 11,5 | 12 | 12,5 | 13 |
| *1* | -26,2 | -25,0 | -26,5 | -25,2 | -23,3 | -21,2 | -17,3 | -15,6 | -11,3 | -9,4 | -6,2 | -2,7 | 1,4 | 5,2 | 9,0 |
| *7* | -26,0 | -25,0 | -26,5 | -25,2 | -25,1 | -21,2 | -19,1 | -15,6 | -13,3 | -9,4 | -7,6 | -4,3 | -1,0 | 3,2 | 7,1 |
| *2* | -25,7 | -26,0 | -26,2 | -25,0 | -23,1 | -22,9 | -21,0 | -17,0 | -14,8 | -9,8 | -7,9 | -3,6 | -1,5 | 2,8 | 5,6 |
| *6* | -25,7 | -26,0 | -26,2 | -25,0 | -21,6 | -22,9 | -19,2 | -17,0 | -12,3 | -9,8 | -5,8 | -3,6 | 1,4 | 2,8 | 8,1 |
| *3* | -25,3 | -25,7 | -26,0 | -26,2 | -23,2 | -21,0 | -19,5 | -17,0 | -14,8 | -11,0 | -8,1 | -2,7 | -0,4 | 2,4 | 4,9 |
| *5* | -25,3 | -25,7 | -26,0 | -26,2 | -23,2 | -21,0 | -19,5 | -17,0 | -12,8 | -11,0 | -5,8 | -2,7 | -0,4 | 2,4 | 7,6 |
| *4* | -24,7 | -25,3 | -25,7 | -26,0 | -23,2 | -18,6 | -15,9 | -13,6 | -11,6 | -9,4 | -7,1 | -4,6 | -2,1 | 2,5 | 5,8 |

**Table 6.** Measured coordinates of the hour markers of the inverted analemmatic sundial along the *Y*-axis: $n$ – the number of the point on the analemma, $y_{n\ meas}$ – the measured coordinates along the *Y*-axis of the $n$-th points on the analemmas of the hour marks in the interval from 06:00 to 13:00, $t$ – the time (hours).

| | \multicolumn{15}{c|}{$y_{n\ meas}$} | | | | | | | | | | | | | | |
|---|---|---|---|---|---|---|---|---|---|---|---|---|---|---|---|
| t\n | 6 | 6,5 | 7 | 7,5 | 8 | 8,5 | 9 | 9,5 | 10 | 10,5 | 11 | 11,5 | 12 | 12,5 | 13 |
| *1* | -1,2 | 2,3 | 5,8 | 8,9 | 11,5 | 12,4 | 14,6 | 14,7 | 16,3 | 16,8 | 17,8 | 18,5 | 19,0 | 18,9 | 17,9 |
| *7* | -1,6 | 2,3 | 5,8 | 8,9 | 10,8 | 12,4 | 13,5 | 14,7 | 15,6 | 16,8 | 17,3 | 18,1 | 18,8 | 18,8 | 18,5 |
| *2* | -3,7 | -1,6 | -1,2 | 2,3 | 4,2 | 7,6 | 9,1 | 10,9 | 11,5 | 13,0 | 13,6 | 13,7 | 13,7 | 14,1 | 13,7 |
| *6* | -3,7 | -1,6 | -1,2 | 2,3 | 6,2 | 7,6 | 10,0 | 10,9 | 11,9 | 13,0 | 13,5 | 13,7 | 14,4 | 14,1 | 13,9 |
| *3* | -5,9 | -3,7 | -1,6 | -1,2 | 1,0 | 3,8 | 6,9 | 8,6 | 8,8 | 10,7 | 11,5 | 11 | 11,5 | 11,9 | 11,6 |
| *5* | -5,9 | -3,7 | -1,6 | -1,2 | 1,0 | 3,8 | 6,9 | 8,6 | 10,0 | 10,7 | 11,7 | 11 | 11,5 | 11,9 | 10,6 |
| *4* | -8,6 | -5,9 | -3,7 | -1,6 | 1,0 | 3,8 | 4,7 | 6,3 | 7,4 | 8,2 | 8,8 | 6,7 | 9,4 | 6,9 | 7,4 |

Knowing the calculated and measured coordinates of hour marks, you can determine how they differ from each other. Tables 7 and 8 show the deviations of the measured from the calculated coordinates of hour marks:

$$\Delta x_n = |x_{n\text{ meas}} - x_n|$$

$$\Delta y_n = |y_{n\text{ meas}} - y_n|$$

**Table 7.** Deviation of the measured from the calculated coordinates of the hour markers of the inverted analemmatic sundial along the *X*-axis: *n* – the number of the analemma point, *Δx_n* – the deviation along the *X*-axis in the interval from 6 to 13 hours, *t* – the time (hours).

| t \ n | \multicolumn{15}{c}{Δx_n} |
|---|---|---|---|---|---|---|---|---|---|---|---|---|---|---|---|
|  | 6 | 6,5 | 7 | 7,5 | 8 | 8,5 | 9 | 9,5 | 10 | 10,5 | 11 | 11,5 | 12 | 12,5 | 13 |
| 1 | 1,6 | 0,6 | 2,7 | 2,5 | 2,1 | 1,8 | 0,1 | 0,9 | 0,6 | 0,5 | 0,4 | 0,2 | 0,6 | 1,1 | 1,6 |
| 2 | 0,2 | 0,3 | 1,1 | 1,0 | 0,6 | 2,2 | 2,5 | 1,0 | 1,6 | 0,4 | 0,8 | 0,2 | 1,0 | 0,0 | 0,5 |
| 3 | 1,2 | 0,6 | 0,3 | 1,6 | 0,1 | 0,3 | 0,4 | 0,4 | 1,0 | 0,2 | 0,4 | 1,7 | 0,7 | 0,2 | 0,6 |
| 4 | 1,4 | 0,6 | 0,4 | 1,8 | 0,5 | 2,3 | 2,8 | 2,6 | 1,8 | 1,0 | 0,2 | 0,6 | 1,4 | 0,1 | 0,1 |
| 5 | 0,1 | 0,7 | 1,6 | 2,9 | 1,4 | 1,0 | 1,7 | 1,7 | 0,3 | 1,5 | 0,6 | 0,4 | 0,6 | 1,1 | 0,8 |
| 6 | 0,5 | 1,0 | 1,8 | 1,7 | 0,2 | 2,9 | 1,4 | 1,7 | 0,2 | 0,3 | 0,6 | 0,5 | 1,2 | 0,7 | 1,3 |
| 7 | 0,2 | 0,6 | 1,5 | 1,3 | 2,7 | 0,6 | 0,7 | 0,3 | 0,2 | 0,7 | 0,6 | 0,6 | 0,6 | 0,3 | 0,9 |

**Table 8.** Deviation of the measured from the calculated coordinates of the hour markers of the inverted analemmatic sundial along the *Y*-axis: *n* – the number of the analemma point, *Δy_n* – the deviation along the *Y*-axis in the interval from 6 to 13 hours, *t* – the time (hours).

| t \ n | \multicolumn{15}{c}{Δy_n} |
|---|---|---|---|---|---|---|---|---|---|---|---|---|---|---|---|
|  | 6 | 6,5 | 7 | 7,5 | 8 | 8,5 | 9 | 9,5 | 10 | 10,5 | 11 | 11,5 | 12 | 12,5 | 13 |
| 1 | 1,2 | 0,0 | 1,1 | 2,0 | 2,5 | 1,4 | 1,9 | 0,4 | 0,3 | 1,2 | 0,0 | 1,1 | 1,2 | 0,9 | 0,5 |
| 2 | 0,2 | 0,4 | 2,4 | 1,1 | 1,3 | 0,1 | 0,1 | 0,1 | 1,6 | 0,9 | 0,7 | 0,2 | 0,6 | 0,4 | 0,2 |
| 3 | 0,7 | 0,6 | 0,3 | 1,5 | 1,4 | 0,6 | 0,8 | 0,9 | 1,2 | 1,7 | 0,3 | 0,2 | 0,3 | 0,5 | 0,8 |
| 4 | 0,7 | 0,3 | 0,5 | 0,6 | 0,1 | 0,7 | 0,1 | 0,1 | 1,3 | 0,5 | 1,1 | 2,8 | 0,5 | 3,2 | 2,1 |
| 5 | 1,1 | 1,0 | 0,7 | 1,1 | 1,0 | 0,2 | 1,2 | 1,3 | 0,4 | 2,1 | 0,9 | 0,6 | 0,7 | 0,9 | 0,2 |
| 6 | 0,3 | 0,1 | 1,9 | 0,6 | 1,2 | 0,6 | 1,3 | 0,6 | 0,7 | 1,4 | 0,3 | 0,3 | 0,6 | 0,1 | 0,5 |
| 7 | 1,2 | 0,4 | 1,5 | 2,4 | 2,2 | 1,8 | 1,2 | 0,8 | 0,6 | 1,6 | 0,1 | 1,1 | 1,4 | 1,2 | 1,5 |

Cup marks on the Belogorsk slab have, basically, a diameter of 1.5 ÷ 2.0 cm, and the distance between two cup marks in one row, corresponding to two adjacent clocks, is on average about 5 cm. That is, the shadow from the gnomon will move directly along each cup mark for about 15÷20 minutes.

In cases where the deviation of the measured coordinates of the hour marks from the calculated coordinates exceeded the average radius of the dimple - about 0.9 cm, the average radius was subtracted from the deviation value. And the remainder was considered as the value of the error in the marking of the cup mark $\Delta x_{n\text{ ed}}$ relative to the calculated hour mark (the distance between the edge of the cup mark and the hour mark corresponding to it). When the deviation did not exceed the average radius of the cup mark, then such a cup mark was considered as coinciding with the hour mark (table 9, 10).

In the case under consideration, it is sufficient to take into account only deviations along the *X*-axis. This is due to the fact that the length of the shadow from the gnomon does not coincide with the distance from the place of its attachment to the corresponding row of cup marks for each month. In order for the shadow from the gnomon to reach the first row of cup marks at the summer solstice, it is necessary that the height of the gnomon will be about 26 cm. Then the length of the shadow will be about 10 cm. With this height of the gnomon, one month before (and one month after) the summer solstice, the length of the shadow will be about 12 cm and the shadow will reach the second row of cup marks. Two months before (and two months after) it will be about 17.5 cm and will immediately reach the fourth row of cup marks. Three months before (and three months after) - on the day of the equinox, the shadow from the gnomon will be about 26 cm and will reach the edge of the plate. That is, the shadow from the gnomon will accurately reach the hour marks only for the first month before and after the summer solstice, and in all other months the length of the shadow will be greater than the value of the coordinates of the hour marks of the corresponding rows along the *Y*-axis. Therefore, to assess the accuracy of measuring time in the case for the Belogorsk plate, it is enough to rely on the error in the marking along the *X*-axis.

**Table 9.** The error of marking the cup marks along the *X*-axis: $n$ – the number of the point on the analemma (the number of the row of cup marks), $\Delta x_{n\ ed}$ – the error of marking the cup marks along the *X*-axis in the interval from 6 to 13 hours, $\Delta x_{n\ ed\ av}$ – the average error of marking the cup marks in the row; t – time (hours).

| t \ n | $\Delta x_{n\ ed}$ | | | | | | | | | | | | | | | $\Delta x_{n\ ed\ av}$ |
|---|---|---|---|---|---|---|---|---|---|---|---|---|---|---|---|---|
| | 6 | 6,5 | 7 | 7,5 | 8 | 8,5 | 9 | 9,5 | 10 | 10,5 | 11 | 11,5 | 12 | 12,5 | 13 | |
| 1 | 0,7 | 0,0 | 1,8 | 1,6 | 1,2 | 0,9 | 0,0 | 0,0 | 0,0 | 0,0 | 0,0 | 0,0 | 0,0 | 0,2 | 0,7 | 0,5 |
| 2 | 0,0 | 0,0 | 0,2 | 0,1 | 0,0 | 1,3 | 1,6 | 0,1 | 0,7 | 0,0 | 0,0 | 0,0 | 0,1 | 0,0 | 0,0 | 0,3 |
| 3 | 0,3 | 0,0 | 0,0 | 0,7 | 0,0 | 0,0 | 0,0 | 0,0 | 0,1 | 0,0 | 0,0 | 0,8 | 0,0 | 0,0 | 0,0 | 0,1 |
| 4 | 0,5 | 0,0 | 0,0 | 0,9 | 0,0 | 1,4 | 1,9 | 1,7 | 0,9 | 0,1 | 0,0 | 0,0 | 0,5 | 0,0 | 0,0 | 0,5 |
| 5 | 0,0 | 0,0 | 0,7 | 2,0 | 0,5 | 0,1 | 0,8 | 0,8 | 0,0 | 0,6 | 0,0 | 0,0 | 0,0 | 0,2 | 0,0 | 0,4 |
| 6 | 0,0 | 0,1 | 0,9 | 0,8 | 0,0 | 2,0 | 0,5 | 0,8 | 0,0 | 0,0 | 0,0 | 0,0 | 0,3 | 0,0 | 0,4 | 0,4 |
| 7 | 0,0 | 0,0 | 0,6 | 0,4 | 1,8 | 0,0 | 0,0 | 0,0 | 0,0 | 0,0 | 0,0 | 0,0 | 0,0 | 0,0 | 0,0 | 0,2 |

**Table 10.** The error of marking the cup marks along the *Y*-axis: $n$ – the number of the point on the analemma (the number of the row of cup marks), $\Delta y_{n\ ed}$ – the error of marking the cup marks along the *Y*-axis in the interval from 6 to 13 hours, $\Delta y_{n\ ed\ av}$ – the average error of marking the cup marks in the row; t – time (hours).

| t \ n | $\Delta y_{n\ ed}$ | | | | | | | | | | | | | | | $\Delta y_{n\ ed\ av}$ |
|---|---|---|---|---|---|---|---|---|---|---|---|---|---|---|---|---|
| | 6 | 6,5 | 7 | 7,5 | 8 | 8,5 | 9 | 9,5 | 10 | 10,5 | 11 | 11,5 | 12 | 12,5 | 13 | |
| 1 | 0,3 | 0,0 | 0,2 | 1,1 | 1,6 | 0,5 | 1,0 | 0,0 | 0,0 | 0,3 | 0,0 | 0,2 | 0,3 | 0,0 | 0,0 | 0,4 |
| 2 | 0,0 | 0,0 | 1,5 | 0,2 | 0,4 | 0,0 | 0,0 | 0,0 | 0,7 | 0,0 | 0,0 | 0,0 | 0,0 | 0,0 | 0,0 | 0,2 |
| 3 | 0,0 | 0,0 | 0,0 | 0,6 | 0,5 | 0,0 | 0,0 | 0,0 | 0,3 | 0,8 | 0,0 | 0,0 | 0,0 | 0,0 | 0,0 | 0,1 |
| 4 | 0,0 | 0,0 | 0,0 | 0,0 | 0,0 | 0,0 | 0,0 | 0,0 | 0,4 | 0,0 | 0,2 | 1,9 | 0,0 | 2,3 | 1,2 | 0,4 |
| 5 | 0,2 | 0,1 | 0,0 | 0,2 | 0,1 | 0,0 | 0,3 | 0,4 | 0,0 | 1,2 | 0,0 | 0,0 | 0,0 | 0,0 | 0,0 | 0,2 |
| 6 | 0,0 | 0,0 | 1,0 | 0,0 | 0,3 | 0,0 | 0,4 | 0,0 | 0,0 | 0,5 | 0,0 | 0,0 | 0,0 | 0,0 | 0,0 | 0,1 |
| 7 | 0,3 | 0,0 | 0,6 | 1,5 | 1,3 | 0,9 | 0,3 | 0,0 | 0,0 | 0,7 | 0,0 | 0,2 | 0,5 | 0,3 | 0,6 | 0,5 |

Thus, on average, the edges of the cup marks in the rows along the X axis are separated from the calculated hour marks by no more than 0.5 cm. This value is equivalent to approximately 5÷6 minutes. That is, the accuracy with which the time was measured by the Belogorsk sundial is about 5÷6 minutes. This is quite good accuracy for a measuring instrument of that era. Thus, the accuracy of time measurement using a Bronze Age cumulative water clock was about 9÷10 minutes (Vodolazhskaya, Usachuk, Nevsky, 2015, 65-87). The higher accuracy of time measurement using sundial compared to water clocks in the Bronze Age in the Northern Black Sea region confirms the hypothesis that the invention of sundial could be associated with the need to improve the accuracy of time measurement.

## References

Mayall, Mayall, 1994 – Mayall, R.N.; Mayall, M.W. Sundials: Their Construction and Use. Cambridge, Mass.: Sky Publishing, 1994.

Petricevic, Vodolazhskaya, 2021 – Petricevic, M.B.; Vodolazhskaya, L.N. Bronze Age sundial from Prokletije (Montenegro). Archaeoastronomy and Ancient Technologies 2021, 9(2), 51-65.

Rohr, 1965 – Rohr, R.R.J. Les Cadrans Solaires. Published by Gauthier-Villars Editeur, 1965.

Savoie, 2009 – Savoie, D. Sundials design construction and use; Published by Springer-Verlag New York Inc., United States, 2009.

Vodolazhskaya, 2013 – Vodolazhskaya L.N. Analemmatic and horizontal sundials of the Bronze Age (Northern Black Sea Coast). Archaeoastronomy and Ancient Technologies 2013, 1(1), 68-88.

Vodolazhskaya, 2022 – Vodolazhskaya L.N. Inverted analemmatic sundial of the Bronze Age. Archaeoastronomy and Ancient Technologies 2022, 10(1), 1-29.

Vodolazhskaya, Larenok, Nevsky, 2014 – Vodolazhskaya L.N., Larenok P.A., Nevsky M.Yu. Ancient astronomical instrument from Srubna burial of kurgan field Tavriya-1 (Northern Black Sea Coast). Archaeoastronomy and Ancient Technologies 2014, 2(2), 31-53.

Vodolazhskaya, Larenok, Nevsky, 2016 – Vodolazhskaya, L.N.; Larenok, P.A.; Nevsky, M.Yu. The prototype of ancient analemmatic sundials (Rostov Oblast, Russia). Archaeoastronomy and Ancient Technologies 2016, 4(1), 96-116.

Vodolazhskaya, Novichikhin, Nevsky, 2021 – Vodolazhskaya, L.N.; Novichikhin, A.M.; Nevsky, M.Yu. Sundial-water clock of the Bronze Age (Northern Black Sea Region). Archaeoastronomy and Ancient Technologies 2021, 9(1), 73-86.

Vodolazhskaya, Usachuk, Nevsky, 2015 – Vodolazhskaya, L.N.; Usachuk, A.N.; Nevsky, M.Yu. Clepsydra of the Bronze Age from the Central Donbass. Archaeoastronomy and Ancient Technologies 2015, 3(1), 65-87.

Waugh, 1973 – Waugh, A.E. Sundials: Their Theory and Construction. New York: Dover Publications, 1973.